\newcommand{\mystyleflag}{0}

\ifnum\mystyleflag=0
\documentclass[12pt]{article}
\pdfoutput=1
\usepackage{jcappub}
\else
\documentclass[journal,article,submit,pdftex,moreauthors]{Definitions/mdpi} 
\firstpage{1} 
\pubvolume{1}
\issuenum{1}
\articlenumber{0}
\pubyear{2024}
\copyrightyear{2024}
\datereceived{ } 
\daterevised{ } 
\dateaccepted{ } 
\datepublished{ } 
\fi

\usepackage{amsmath}
\usepackage{amssymb}
\usepackage{physics}
\usepackage{xcolor}
\usepackage{slashed}
\usepackage{subfigure}
\usepackage{float}
\usepackage{notoccite}
\def\be{\begin{equation}}
\def\ee{\end{equation}}
\def\bea{\begin{eqnarray}}
\def\eea{\end{eqnarray}}

\ifnum\mystyleflag=0
\title{Quasi-equilibrium states and phase transitions in biological evolution}
\author[1]{Artem Romanenko}
\emailAdd{artemius.the.roman@gmail.com}\author[1,2]{and Vitaly Vanchurin} 
\emailAdd{vitaly.vanchurin@gmail.com}
\affiliation[1]{Artificial Neural Computing, Weston, Florida, 33332, USA}
\affiliation[2]{Duluth Institute for Advanced Study, Duluth, Minnesota, 55804, USA}
\begin{document}   
\else
\Title{Quasi-equilibrium states and phase transitions in biological evolution}
\TitleCitation{Quasi-equilibrium states and phase transitions in biological evolution}
\Author{Artem Romanenko $^{1}$ and Vitaly  Vanchurin $^{1,2}$}
\AuthorNames{Artem Romanenko and Vitaly  Vanchurin}
\AuthorCitation{Romanenko, A.; Vanchurin, V.}
\address{%
$^{1}$ \quad Artificial Neural Computing, Weston, Florida, 33332, USA\\
$^{2}$ \quad Duluth Institute for Advanced Study, Duluth, Minnesota, 55804, USA}
\fi

\abstract{We develop a macroscopic description of the evolutionary dynamics by following the temporal dynamics of the total Shannon entropy of sequences, denoted by $S$, and the average Hamming distance between them, denoted by $H$. We argue that a biological system can persist in the so-called quasi-equilibrium state for an extended period, characterized by strong correlations between $S$ and $H$, before undergoing a phase transition to another quasi-equilibrium state. To demonstrate the results, we conducted a statistical analysis of SARS-CoV-2 data from the United Kingdom during the period between March, 2020 and December, 2023. From a purely theoretical perspective, this allows us to systematically study various types of phase transitions described by a discontinuous change in the thermodynamic parameters. From a more practical point of view, the analysis can be used, for example, as an early warning system for pandemics.}

\ifnum\mystyleflag=0
\maketitle  
\else
\begin{document}   
\fi

\section{Introduction}

In molecular biology and population genetics, the biological evolution of biopolymers and viruses is usually described using concepts such as quasispecies \cite{quasi}, fitness landscape \cite{landscape}, error threshold \cite{error}, etc. These concepts first appeared and were applied in the context of RNA models \cite{schuster}, where a genotype-to-phenotype mapping was investigated. A typical application of such a framework to experimental data can be seen in Ref. \cite{koelle}, where Kimura’s neutral theory \cite{kimura}, genotype-phenotype mapping \cite{schuster}, and fitness landscapes \cite{landscape} are utilized.

Quasispecies are usually defined as a population of macromolecular species with closely interrelated sequences, dominated by one or several degenerate master copies \cite{quasi}. This is a useful concept which, as we shall see below, is closely related to a quasi-equilibrium state that can be defined more rigorously using the tools of statistical physics. Roughly speaking, quasi-equilibrium states are  formed by random genetic drift acting on neutral or nearly neutral mutations, rather than by positive selection for advantageous traits, in agreement with the neutral theory \cite{kimura}. In terms of statistical physics, quasi-equilibrium states represent a micro-canonical-type ensemble, with the negative logarithm of fitness playing the role of energy \cite{sella, Vanchurin2}

Although the quasi-equilibrium states are relatively stable and their macroscopic parameters change continuously, sudden discontinuous changes may also occur, leading to a phase transition. One example of the phase transition is the so-called error threshold, a point at which the mutation rate of the replicating molecules surpasses a critical threshold \cite{error}. Below this threshold, the population can maintain a certain level of genetic information, and natural selection can effectively act to preserve functional sequences. Above the threshold, the error rate becomes too high, leading to a loss of information due to the accumulation of deleterious mutations, and the population enters a state of error catastrophe. These phase transitions were recently discussed in Ref. \cite{sole}, where the authors point out the benefit of using simple physical models. In this article, we shall perform a statistical analysis of the quasi-equilibrium states and the phase transitions between such states by following the temporal dynamics of the macroscopic parameters, such as the total entropy and average Hamming distance, using the tools of statistical physics.

Numerous efforts were undertaken to formulate a systematic physical description of biological evolution starting from the early part of the 20th century (see Ref. \cite{wright}). It was widely believed that the laws of biology, particularly those of evolutionary biology, can be somehow derived from the laws of physics, especially statistical physics. Notable attempts to develop a physical theory of biology include renowned works such as Schrödinger's ``What is Life? The Physical Aspect of the Living Cell'' \cite{Schroedinger} and Fisher's ``The Genetical Theory of Natural Selection'' \cite{genetical_theory}. Some approaches involved utilizing well-known physical effects, e.g. resonance \cite{resonance}, to describe evolution and there was even an attempt to apply the formalism of quantum mechanics \cite{quantum}. However, until recently, clear connections between physical or thermodynamic quantities and observable biological parameters had not been established, despite the practice of measuring parameters like enthalpy, entropy, and Gibbs free energy for microorganisms \cite{stockar,popovic,popovic2}. 

The situation began to change at the beginning of the 21st century when Sella and Hirsh \cite{sella} identified the effective population size with inverse temperature, and a few years later Barton and Coe \cite{barton} identified the volume of allele frequency space with entropy, and Pan and Deem \cite{pan} used Shannon entropy to measure diversity and selection pressure for H3N2 influenza sequences. Another important work was by Jones and collaborators \cite{jones} who modeled viral evolution using a grand-canonical-type ensemble. All of these results marked a paradigm shift in the phenomenological description of biological evolution, but if one wishes to derive the description from first principles the starting point should not be the identification of macroscopic parameters, but specification of a statistical ensembles, i.e. a probability distribution over sequences. Then, for example, the entropy would have to be defined as the Shannon entropy of the distribution and the inverse temperature as a Lagrange multiplier which imposes a constraint on the average energy-like quantity (e.g. logarithmic fitness). 

The first-principles approach was recently employed in Ref. \cite{TOL}, where a fully statistical description of biological evolution (modeled as multilevel learning \cite{TTOLaML}) was developed using a theory of machine learning (established earlier in Refs. \cite{Vanchurin1, Vanchurin2}). The primary new thermodynamic concept that accompanies learning, and consequently biological evolution, is that, in addition to the increase in entropy (according to the second law of thermodynamics), the entropy of a learning system may also decrease (according to the second law of learning \cite{Vanchurin1, Vanchurin2}). As we shall argue, this is precisely what happens in biological systems --- entropy increases in the so-called quasi-equilibrium states and decreases after phase transitions. This finding further strengthens the idea that biological evolution can be effectively modeled through learning dynamics, opening up possibilities for investigating various biological phenomena using the framework provided by the theory of learning. Indeed, numerous non-trivial emergent physical phenomena, including quantum mechanics \cite{Vanchurin2, Katsnelson} and gravity \cite{Vanchurin2, Vanchurin3}, as well as critical phenomena such as  phase transitions \cite{Vanchurin1} or scale invariance \cite{SOC}, have already been derived from learning dynamics. This paper, along with Refs. \cite{TTOLaML, TOL}, can be regarded as a step toward modeling emergent biological phenomena within the same mathematical framework of the learning theory \cite{Vanchurin1}. 

The major transitions in evolution, a concept in evolutionary biology popularized by Smith and Szathmary in their book ``The Major Transitions in Evolution'' \cite{major_transitions,majtranstwo}, can be correlated with learning phase transitions as proposed in Ref. \cite{TOL}. Specifically, the `origin-of-life' phase transition has been demonstrated to represent a shift from a grand-canonical ensemble of molecules to a grand-canonical ensemble of organisms. However, in line with the theory of learning \cite{Vanchurin1}, learning phase transitions are a lot more common, and it is not immediately clear how to consistently identify and describe them in the biological context. (See however Refs. \cite{Eigen, Wallace} for other attempts to describe transitions in biological evolution as thermodynamic phase transitions.) For instance, in biological evolution, can the formation of new genetic variants (or formation of new levels \cite{TTOLaML}) be interpreted as a learning phase transition, or what one might call ``minor transition in evolution''? This paper addresses these minor transitions by analyzing the dynamics of macroscopic quantities such as the entropy of sequences, denoted as $S$ (defined as Shannon entropy), and the Hamming distance between sequences, denoted as $H$ (averaged over all pairs of sequences in a given ensemble). It is shown that these minor transitions occur when both the entropy and the Hamming distances undergo a sudden, discontinuous jump. The term \textit{phase transition} comes from statistical physics and can be related to a discontinuous change of a statistical ensemble descried, for example, by free energy. In biology, and particularly in genetics, such a term is not widely used. A more familiar terms is the so-called ``selective sweep'' which occurs when an advantageous mutation gets fixed (along with some other nearby genes on the same allele which is known as ``hitchhiking'' \cite{hitchhiking}). Though selective sweep is an example of a phase transition, there are more general phase transitions that shall be discussed in the paper.

To illustrate the procedure we have performed statistical analysis of ensembles of sequences from SARS-CoV-2 datasets. The main reason for using the SARS-CoV-2 is that the datasets contain an unprecedented amount of data collected during a sufficiently long period of time which allows us to observe and analyze a number of quasi-equilibrium states and phase transitions. Furthermore, the preformed analysis revealed a possibility of using the procedure for developing an early warning system for pandemics which by itself may have an independent value. 

The paper is organized as follows. In Sec. \ref{Sec:Ensembles}, we describe the datasets and define statistical ensembles. In Secs. \ref{Sec:Entropy} and \ref{Sec:Hamming}, we approximate respectively the total Shannon entropy of the system and the average Hamming distances for the statistical ensembles. In Sec. \ref{Sec:Equilibrium}, we identify and analyze the quasi-equilibrium states and phase transitions between them. In Sec. \ref{Sec:Conclusion}, we discuss the main results of the paper.

\section{Statistical ensembles}\label{Sec:Ensembles}

A statistical ensemble is typically defined as a probability distribution over configuration space (or phase space). In the context of biological evolution, the relevant configuration space is the space of genomes, known as the genotype space. Individual points in the genotype space represent genome sequences, drawn from an alphabet of four characters or nucleobases: adenine (A), cytosine (C), guanine (G), and thymine (T). Considering sequences of a fixed length $K$ and each site having one of four states, there are $4^K$ distinct points (or states) in the relevant genotype space. Not all of these states are equally probable, and to define a probability distribution, one can either model it analytically or extract it from experimental data. In this paper, we employ the experimental approach to study the statistical properties of an evolving biological system, such as the coronavirus.

For the numerical analysis, genome sequence data were extracted from the NCBI Virus SARS-CoV-2 Data Hub, the GenBank \cite{ncbi_virus,genbank}. Specifically, we selected complete genomes collected in the United Kingdom, retaining only those with fewer than 1\% ambiguous characters --- ensuring relatively clean sequences. These sequences were then grouped into statistical ensembles representing different months starting from March 2020, as the first two months of the pandemic (January and February 2020) had too few sequences. Genome sequences from all other months were randomly selected (ranging from at least 1500 to at most 2000 sequences) and aligned using the FFT-NS-2 method implemented in the MAFFT \cite{mafft_website,mafft}. After alignment, probabilities of the nucleobases A, T, G, and C were calculated for each site $i$, i.e.
\bea
p_i^A \equiv p(x_i= A);\;\; p_i^T \equiv p(x_i= T);\;\; p_i^G \equiv p(x_i= G);\;\; p_i^C \equiv p(x_i= C),
\eea
with the appropriate normalization $p_i^A +p_i^T +p_i^G +p_i^C = 1$. These probabilities were then used to replace any absent nucleobases (i.e., gaps or ambiguous characters) with randomly selected nucleobases. Alternatively, we could have introduced an additional `character' for insertions and deletions, but our analysis indicates that this does not significantly modify the macroscopic statistical properties of the ensembles.

For each statistical ensemble, or equivalently, for each probability distribution, $p(\vec{x})$, over sequences $\vec{x} \in \{A, T, G, C\}^K$, two relevant macroscopic quantities were computed: the total Shannon entropy of sequences
\be
S = - \sum_{\vec{x}\in \{A, T, G, C\}^K} p(\vec{x}) \log \left ( p(\vec{x})\right ) \label{Eq:Entropy}
\ee
(see Sec. \ref{Sec:Entropy}) and the average Hamming distance between sequences
\be
H = \sum_{\vec{x},\vec{y}\in \{A, T, G, C\}^K} \; h(\vec{x},\vec{y}) \; p(\vec{x}) \;p(\vec{y}),
\label{Eq:Hamming}
\ee
(see Sec. \ref{Sec:Hamming}) where the Hamming distance between individual sequences is defined as 
\be
h(\vec{x},\vec{y})  = \sum_{i=1}^N \left (1-\delta(x_i, y_i)\right)\label{Eq:h(x,y)}
\ee
and
\be
\delta(a, b) = \begin{cases} 1\;\;\; \text{if}\;\;\; a=b \\
0 \;\;\;\text{if} \;\;\;a\neq b
\end{cases}
\ee 
is similar to the Kronecker delta function. In addition, we calculated the average Hamming distances from individual sequences, i.e. the average distance to all other sequences in the ensemble. The fractional part of this distribution, as discussed in Sec. \ref{Sec:Hamming}, contains some nontrivial information about the ensembles, or, more precisely, about the network of neutral mutations\cite{frank1,frank2,neutral_mutations}. 

One of the main objectives in our studies is the analysis of the temporal dynamics of the statistical ensembles, described by the macroscopic quantities $S$ and $H$, over extended periods of time. This allowed us to identify the so-called quasi-equilibrium states (usually lasting a few months) during which an approximate linear dependence holds,
\be 
S \approx a H + b.\label{Eq:SH}
\ee
Phase transitions between quasi-equilibrium states occur when the parameters $a$ and $b$ change discontinuously (see Sec. \ref{Sec:Equilibrium}).

\section{Entropy}\label{Sec:Entropy}

For starters, consider an ensemble of sequences with the average single-site Shannon entropy defined as
\bea
S_1 &=& - \frac{1}{K} \sum_{i=1}^{K}\;\;\; \sum_{x_i \in \{A, T, G, C\}}  p(x_i)  \log \left ( p(x_i) \right ) \label{Eq:S1} \\ \notag
&=& - \frac{1}{K} \sum_{i=1}^{K} \left ( p_i^A \log \left ( p_i^A \right ) +p_i^T \log \left ( p_i^T \right )+p_i^G \log \left ( p_i^G \right )+p_i^C \log \left ( p_i^C \right ) \right ).
\eea
The above equation involves averaging over both: an ensemble of $N$ sequences (the inner summation) and over $K$ sites with non-zero entropy (the outer summation). We will refer to the sites with non-zero entropy as `active' sites and the sites with zero entropy (i.e., with a unique nucleobase) as `passive' sites. In what follows, passive sites will be excluded from the analysis since they do not contribute to the statistical properties of the system.

The single-site entropy \eqref{Eq:S1} can be fairly accurately estimated by analyzing a finite number of sequences, or samples, from the ensemble. Unfortunately, the analysis breaks down if we attempt the same `brute-force' approach to calculate the total entropy of the system, of all active sites \eqref{Eq:Entropy}. The problem arises because the number of possible states grows exponentially with the number of sites $K$, and if the number of sequences is fixed (due to computational constraints, e.g. $N=2000$), then we cannot reliably estimate a probability distribution for more than $\log (N) = \log (2000) \sim 7.6$ sites.

To overcome these computational constraints, we first calculate entropies for only a few consecutive sites $S_1, S_2, S_3, \ldots$ and then extrapolate to $S_K$. In fact, it is more convenient to extrapolate the entropy per site $S_1/1, S_2/2, S_3/3, \ldots$ which usually scales as a decaying exponential, i.e.,
\be
\frac{S_m}{m} = A \exp(- B m) + C,\label{Eq:Exponential}
\ee
for constants $A$, $B$ and $C$ which may vary between ensembles, but are assumed to be fixed within a given ensemble. The average entropy per site can be estimated as 
\be
    \frac{S_m}{m} =- \frac{\sum_{i=1}^{K-m+1}\;\;\; \sum_{(x_i, ..., x_{i+m-1}) \in \{A, T, G, C\}^m}  p(x_i, ..., x_{i+m-1})  \log \left ( p(x_i, ..., x_{i+m-1}) \right )}{m(K-m+1)}\label{Eq:Sk}
\ee
where the averaging is over both: $N$ samples (the inner summation) and $K-m+1$ choices of consecutive sites (the outer summation). 

\begin{figure}
    \centering
    \includegraphics[width=0.8\textwidth]{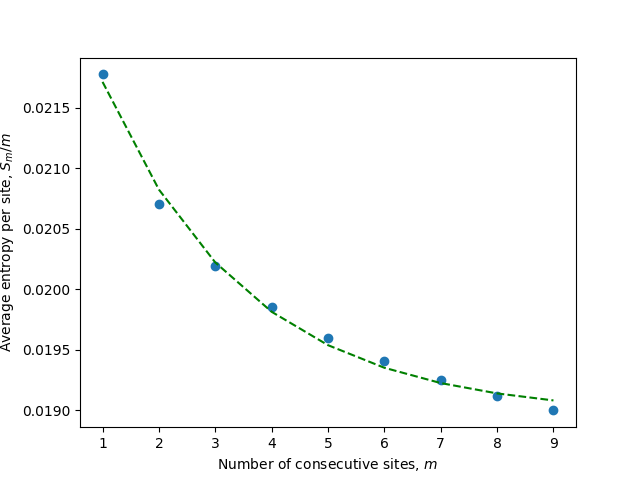}
    \caption{Entropy per site as a function of the number of consecutive sites for April, 2022.}
    \label{fig:extrapolation_good}
\end{figure}

It is important to emphasize once again that the entropy is calculated for only active consecutive sites (after eliminating passive sites with zero entropy). These are the only sites contributing to the statistical properties of the system and, at the same time, may include non-trivial correlations between neighboring sites. Indeed, it is well-known that local correlations in the sequences are much stronger than non-local correlations. For example, three consecutive nucleotides (a trinucleotide) encode a single amino acid, and because of that, there are stronger correlations between sites at distances of order three or less. This phenomena can be observed on Fig. \ref{fig:extrapolation_good}, where parameters of an exponential fit of Eq. \eqref{Eq:Exponential} are estimated by considering different numbers of consecutive sites \eqref{Eq:Sk}. In Table \ref{tab:extrapolation_parameters}, we have compiled the parameters from Eq. \eqref{Eq:Exponential} for all ensembles corresponding to different months. It is evident that the exponential decay rate $B$ is approximately on the order of $1/3$, in accordance with our expectations.

\begin{table}[]
    \centering
    \begin{tabular}{|c|c|c|c|c|c|c|c|c|c|}
        \hline
        N & Month & $A \cdot 10^3$ & $B \cdot 10$ & $C \cdot 10^2$ & N & Month & $A \cdot 10^3$ & $B \cdot 10$ & $C \cdot 10^2$ \\
        \hline
        3 & 2020-03 & 2.49 & 5.56 & 1.95 & 26 & 2022-02 & 2.73 & 3.04 & 2.16 \\
        4 & 2020-04 & 1.83 & 6.51 & 1.52 & 27 & 2022-03 & 1.78 & 2.07 & 1.82 \\
        5 & 2020-05 & 1.68 & 6.14 & 1.77 & 28 & 2022-04 & 1.15 & 5.21 & 1.33 \\
        6 & 2020-06 & 1.19 & 4.76 & 2.10 & 29 & 2022-05 & 0.73 & 2.37 & 1.54 \\
        7 & 2020-07 & 1.63 & 4.94 & 2.88 & 30 & 2022-06 & 1.48 & 3.77 & 1.95 \\
        8 & 2020-08 & 2.48 & 4.50 & 3.16 & 31 & 2022-07 & 0.98 & 4.28 & 1.76 \\
        9 & 2020-09 & 2.12 & 3.87 & 2.97 & 32 & 2022-08 & 0.78 & 2.76 & 1.75 \\
        10 & 2020-10 & 1.63 & 4.85 & 2.53 & 33 & 2022-09 & 1.09 & 1.60 & 1.99 \\
        11 & 2020-11 & 1.63 & 4.66 & 2.39 & 34 & 2022-10 & 1.26 & 1.27 & 2.11 \\
        12 & 2020-12 & 2.44 & 5.02 & 2.51 & 35 & 2022-11 & 1.70 & 1.43 & 2.30 \\
        13 & 2021-01 & 1.59 & 4.71 & 1.91 & 36 & 2022-12 & 1.59 & 2.33 & 2.44 \\
        14 & 2021-02 & 1.10 & 4.96 & 1.60 & 37 & 2023-01 & 2.25 & 2.22 & 2.73 \\
        15 & 2021-03 & 0.91 & 3.82 & 2.01 & 38 & 2023-02 & 2.80 & 2.41 & 2.61 \\
        16 & 2021-04 & 1.55 & 3.76 & 2.48 & 39 & 2023-03 & 1.96 & 2.44 & 2.25 \\
        17 & 2021-05 & 4.12 & 2.50 & 4.04 & 40 & 2023-04 & 1.43 & 1.16 & 1.88 \\
        18 & 2021-06 & 2.64 & 2.09 & 2.84 & 41 & 2023-05 & 1.05 & 1.56 & 2.10 \\
        19 & 2021-07 & 0.62 & 2.53 & 1.68 & 42 & 2023-06 & 1.76 & 1.45 & 3.16 \\
        20 & 2021-08 & 0.55 & 2.10 & 1.71 & 43 & 2023-07 & 1.73 & 1.11 & 2.94 \\
        21 & 2021-09 & 0.73 & 3.86 & 1.59 & 44 & 2023-08 & 1.06 & 2.03 & 2.55 \\
        22 & 2021-10 & 0.32 & 2.42 & 1.69 & 45 & 2023-09 & 1.13 & 1.83 & 2.52 \\
        23 & 2021-11 & 0.68 & 4.10 & 1.73 & 46 & 2023-10 & 1.80 & 1.38 & 2.66 \\
        24 & 2021-12 & 4.31 & 2.06 & 2.65 & 47 & 2023-11 & 4.28 & 1.35 & 3.25 \\
        25 & 2022-01 & 4.06 & 3.90 & 1.89 & 48 & 2023-12 & 6.69 & 2.05 & 3.33 \\
        \hline
    \end{tabular}
    \caption{Parameters of the entropy per site in Eq. \ref{Eq:Exponential} for different months.}
    \label{tab:extrapolation_parameters}
\end{table}

\begin{figure}
    \centering
    \includegraphics[width=0.9\textwidth]{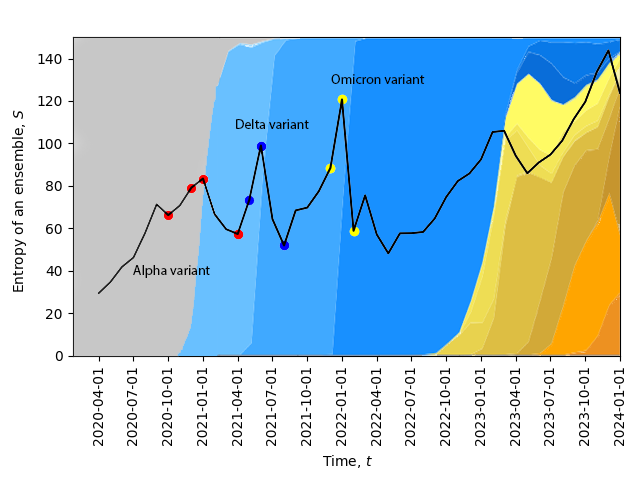}
    \caption{Evolution of the total entropy with an overlay of the percentage of cases attributed to different variants according to the GISAID website\cite{gisaid_website,gisaid}.}
    \label{fig:entropy_with_covid}
\end{figure}

Essentially the same analysis was performed for all ensembles (i.e., all months), and then the total entropy of each ensemble was calculated,
\be
S  = S_K  = K \frac{S_K}{K} \approx K \left (A \exp(-B k) + C \right ) \approx K C. \label{Eq:S}
\ee
On Fig. \ref{fig:entropy_with_covid} we plot evolution of the total entropy \eqref{Eq:S} as a function of time. The growth of the total entropy indicates that the virus spreads across the space of neutral mutations, or what we call the neutral network. This high-entropy state corresponds to a higher diversity of genomes, allowing for a larger volume of genotype space to be explored for possible positive mutations (i.e., mutations to states with higher fitness). The entropy growth accelerates further when the positive mutation is found, and then the system quickly undergoes a phase transition. After this transition, the entropy drops, corresponding to a new variant replacing the old one and in agreement with the second law of learning \cite{Vanchurin1, TOL}. On Fig. \ref{fig:entropy_with_covid}, we can already identify four phase transitions at December 2020, May 2021, December 2021 and November 2023, and additional phase transitions will be identified in Sec. \ref{Sec:Equilibrium}. Note that the peak at November 2023, suggests that the system might be currently undergoing a phase transition. 

The temporal dynamic of the total entropy of Fig. \ref{fig:entropy_with_covid} can be compared with data representing the appearance of different variants of the virus. On Fig. \ref{fig:entropy_with_covid} the percentage of cases relative to different virus variants form Ref. \cite{gisaid} was added on the background. The four red dots represent: the first case of Alpha variant in the UK, 5\% of all cases, 50\% of all cases, 95\% of all cases; the three dark blue dots represent: Delta variant 5\% of all cases, 50\% of all cases, 95\% of all cases; and the three yellow dots represent: Omicron variant 5\% of all cases, 50\% of all cases, 95\% of all cases. Evidently, the replacement of the old variant with a new one is a phase transition which is accompanied with a sudden increase in the total entropy, and after the phase transition the total entropy drops as expected. For example, when the Delta variant started to displace the older Alpha variant or when the Omicron variant started to displace the Delta variant we clearly see sharp peaks of the total entropy. On the far right part of the figure the entropy had gradually increased for a long period of time along with the appearance of multiple new variants, which may indicate a multi-level quasi-equilibrium state to be discussed below in Sec. \ref{Sec:Equilibrium}.

\section{Hamming distance}\label{Sec:Hamming}

\begin{figure}
    \centering
    \includegraphics[width=1.0\textwidth]{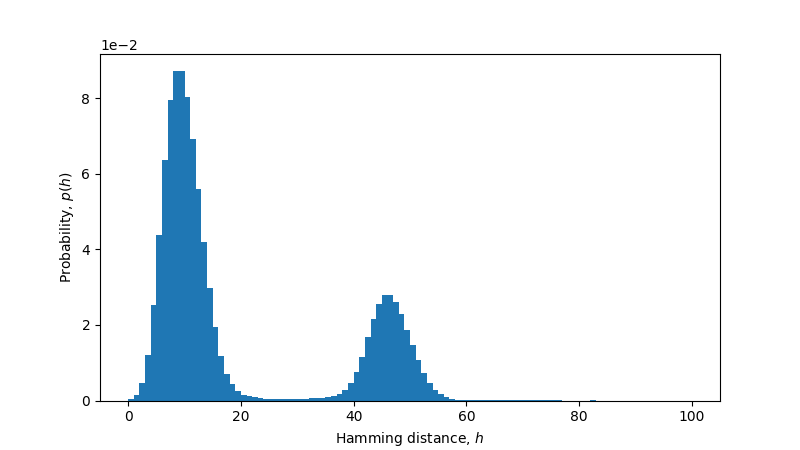}
    \caption{Distribution of Hamming distances for January of 2021.}
    \label{fig:spectrum}
\end{figure}

A natural measure of distances in the genotype space is the Hamming distance, which is also a common metric for comparing sequences of letters. In our case, these sequences consist of nucleobases A, T, G and C, and the Hamming distance between such sequences is defined as the number of sites with different nucleobases (see Eq. \eqref{Eq:h(x,y)}). The measure is well-defined for sequences with no ambiguous characters, but to generalize this measure to all aligned sequences and thus define statistical ensembles in the genotype space, we substituted ambiguous characters with random characters drawn from a marginal probability distribution for the corresponding site (see Sec. \ref{Sec:Ensembles}). On Fig. \ref{fig:spectrum} we plot an example of the probability distribution of Hamming distances $p(h)$ between random pairs of sequences, where the Hamming distances $h(\vec{x},\vec{y})$ are treated as random variables. The two peaks represent the presence of at least two variants in the virus populations, as discussed below.

\begin{figure}
    \centering
    \includegraphics[width=1.0\textwidth]{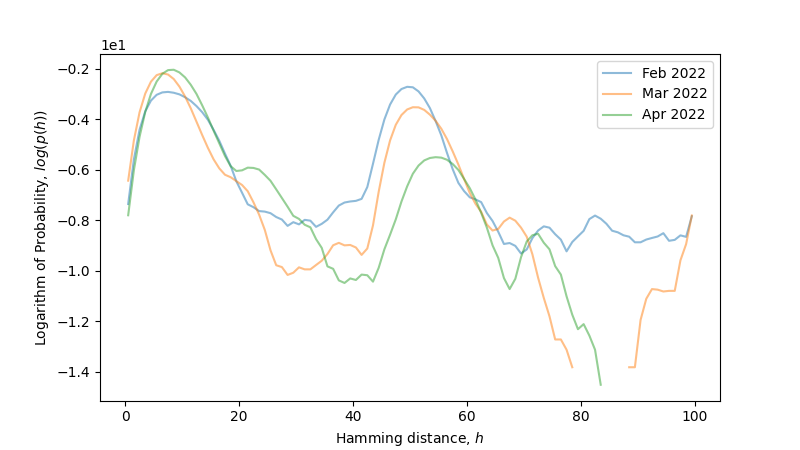}
    \caption{Evolution of distribution of Hamming distances after a phase transition from February to April of 2022.}
    \label{fig:log_death_of_a_peak}
\end{figure}

Assuming that the evolving system is in a local equilibrium state, individual sequences might undergo mutations, but the overall distribution of sequences $p(\vec{x})$ as well as the distribution of Hamming distances $p(h)$ would remain constant. For example, the formation of a local equilibrium state after a phase transition can be observed in Fig. \ref{fig:log_death_of_a_peak}
. From February to April 2022, the rightmost peak consistently decreases (a remnant of the old phase), while the leftmost peak steadily increases (describing the new phase). Note that the logarithm of the distribution is plotted on Fig. \ref{fig:log_death_of_a_peak} as opposed to Fig. \ref{fig:spectrum}, providing more information for distributions with exponential tails, like exponential or normal distributions.

In general, the presence of multiple peaks corresponds to different characteristic scales, such as the mean distance within a cluster, mean distance between clusters, mean distances between clusters of clusters, etc. If the additional peaks remain fixed over extended periods of time, then this would indicate that the evolving system has developed new scales (or levels) and can thus be described as a multi-level learning system \cite{TTOLaML, TOL}. There may also be more fine-grained details of the quasi-equilibrium states that are not captured by the analysis of the total entropy or average Hamming distances but can be extracted only by considering the spectral properties of the neutral network, i.e., the network of neutral mutations. We leave the analysis of the spectral properties for future research.

\begin{figure}
    \centering
    \includegraphics[width=1.0\textwidth]{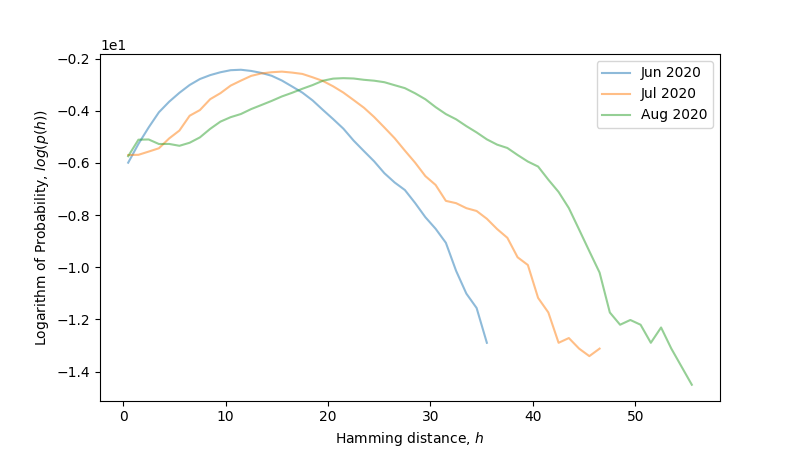}
    \caption{Evolution of distribution of Hamming distances in a quasi-equilibrium state from June to August of 2020.}
    \label{fig:moving_peak}
\end{figure}

Returning to the concept of simple single-level equilibrium states, characterized by a single primary peak (or scale) in the distribution of Hamming distances. If the primary peak is fixed, as in Fig. \ref{fig:log_death_of_a_peak}, then the local genotype space has already been explored, and the system is in a local equilibrium state. However, the general tendency of the system is to explore the genotype space through neutral mutations, in which case the peak should move to larger scales, i.e., to the right. On Fig. \ref{fig:moving_peak}, the system is observed from June till August of 2020, during which time the local neutral network is explored and what we call a quasi-equilibrium state is established.

Given the probability distribution $p(h)$, we can calculate the average Hamming distance \eqref{Eq:Hamming}:
\be
H = \int dh\; h\; p(h).
\ee
Roughly speaking, $H$ tells us about the dispersion of the sequence distribution or the divergence between different genomes. More precisely, it is a combination of multiple effects, dependent on both inter-cluster distances and the average distance within each cluster. Therefore, a large $H$ can be a sign of multiple clusters existing at the same time, for example, during phase transitions between quasi-equilibrium states or, possibly, in a multi-level quasi-equilibrium state. 

Another interesting property of the ensembles can be observed in the distribution of fractional parts of a single-sequence average Hamming distances. In other words, if we define the average Hamming distance from a given sequence $\vec{x}$, i.e.:
\be
f(\vec{x}) = \sum_{\vec{y}\in 4^K} \; h(\vec{x},\vec{y})\;p(\vec{y}),
\ee
and then consider its fractional part $f(\vec{x}) - \lfloor f(\vec{x}) \rfloor$ as a random variable, the corresponding probability distribution $p(f - \lfloor f \rfloor)$ has a rather peculiar form. Indeed, if the space of neutral mutations were filled uniformly, we would also expect to see a uniform distribution of the fractional parts, i.e., $p(f - \lfloor f \rfloor) \approx const$, which is exactly what was observed during phase transitions, but not in the quasi-equilibrium states.

\begin{figure}
    \centering
    \includegraphics[width=1.0\textwidth]{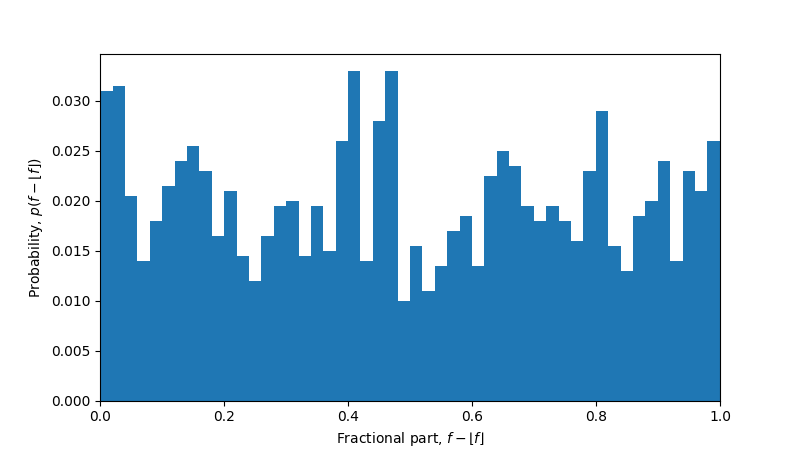}
    \caption{Normalized distribution of fractional parts of single-sequence average Hamming distances from October of 2023.}
    \label{fig:frac_regime_1}
\end{figure}

\begin{figure}
    \centering
    \includegraphics[width=1.0\textwidth]{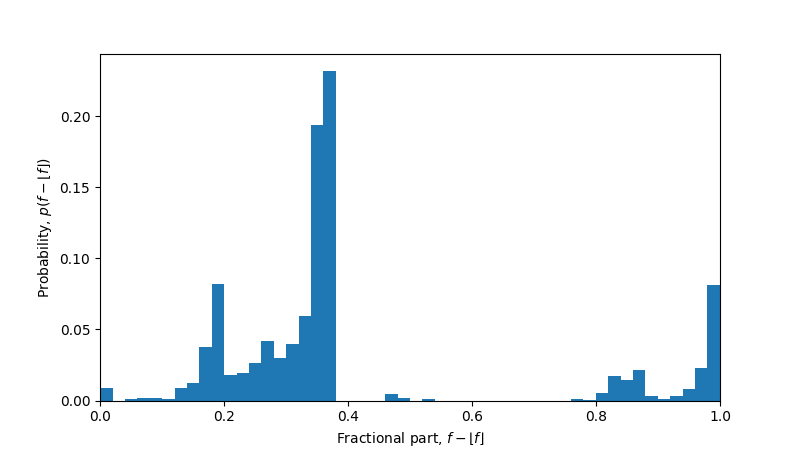}
    \caption{Normalized distribution of fractional parts of single-sequence average Hamming distances from April of 2022.}
    \label{fig:frac_regime_2}
\end{figure}

For example, in October 2023, the system was undergoing a phase transition, and the distribution is nearly uniform, see Fig. \ref{fig:frac_regime_1}, but in April 2022, the system is in a quasi-equilibrium state, and the distribution has a very sharp peak, see Fig. \ref{fig:frac_regime_2}
, i.e. one (or few) values of the fractional part dominates over others values. The peaked distribution indicates that there is a central sequence $\vec{x_0}$, most of the shortest Hamming distances between sequences $\vec{x}$ and $\vec{y}$ go trough that sequence, i.e. 
\be
h(\vec{x}, \vec{y}) \approx h(\vec{x}, \vec{x_0}) +h(\vec{x_0}, \vec{y}) 
\ee
and then the fractional part of all average Hamming distances is approximately the same
\be
f(\vec{x}) - \lfloor f(\vec{x}) \rfloor \approx  f(\vec{x}_0) - \lfloor f(\vec{x_0}) \rfloor 
\ee
for all $\vec{x}$. This may be expected in a single-level quasi-equilibrium state with one or few central sequences, but not as much during phase transitions between quasi-equilibrium states or in a multi-level quasi-equilibrium state.

\section{Quasi-equilibrium states}\label{Sec:Equilibrium}

During phase transitions a system is transferred from one quasi-equilibrium state to another which is accompanied with discontinuous change in the statistical ensembles, which can often be observed as a discontinuous change of macroscopic parameters. In our analysis the two main macroscopic parameters are the total Shannon entropy $S$ (see Sec. \ref{Sec:Entropy}) and the average Hamming distance $H$ (see Sec. \ref{Sec:Hamming}). On the scatter plot in Fig. \ref{fig:p(S,H)_lines} \begin{figure}
    \centering
    \includegraphics[width=1.0\textwidth]{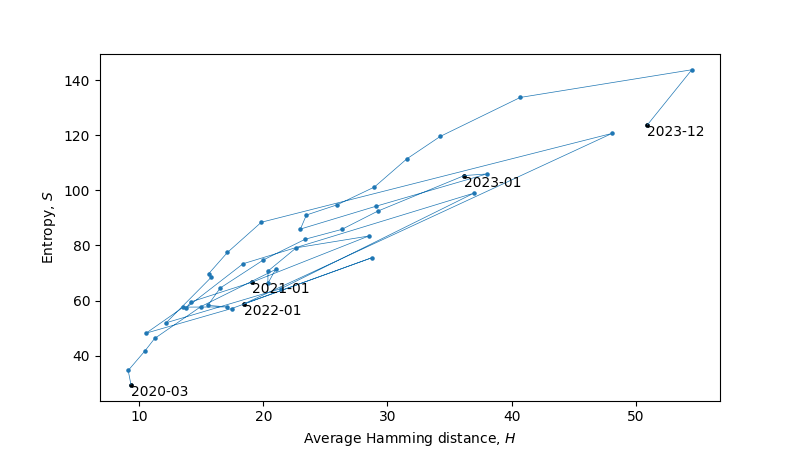}
    \caption{Scatter plot of the total entropy and average Hamming distances with consecutive months connected by lines.}
    \label{fig:p(S,H)_lines} 
\end{figure}
each point represents a single month in the evolution of coronaviruses and consecutive months are connected by lines. In what follows, we will describe a procedure for grouping the consecutive points into a quasi-equilibrium state which can be described by a joint probability distribution $p(S, H)$.

When the system is in a quasi-equilibrium state (i.e., in-between phase transitions), both $S$ and $H$ can change more or less continuously, but an approximately linear dependence between them remains fixed. In a Gaussian limit, such distributions are given by
\be
    p(S,H) \propto \exp\left (-\frac{(S - a H - b)^{2}}{2 \sigma_\perp^2} - \frac{(S + a^{-1} H - c)^{2}}{2 \sigma_\parallel^2} \right ) \label{Eq:pSH}
\ee
with variances $\sigma_\perp \ll \sigma_\parallel$. There may also be non-Gaussianities, the analysis of which is beyond the scope of this paper. In relation to the coronavirus data, we identified eight quasi-equilibrium states whose parameters are summarized in Table \ref{tab:p(S,H)}. 
\begin{table}[]
    \centering
    \begin{tabular}{|c|c|c|c|c|c|}
        \hline
        Duration & a & b & c & $\sigma_\perp^2$ & $\sigma_\parallel^2$ \\
        \hline
        2020-04 to 2020-08 & 2.97 & 10.63 & 54.82 & 4.91 & 204.89 \\
        2020-10 to 2020-11 & 3.71 & -4.92 & 80.73 & 0 & 20.46 \\
        2020-12 to 2021-03 & 1.72 & 34.15 & 77.67 & 0.33 & 186.83 \\
        2021-07 to 2021-11 & 4.81 & -6.30 & 74.55 & 1.08 & 156.01 \\
        2022-01 to 2022-04 & 1.52 & 31.31 & 72.28 & 0.44 & 200.24 \\
        2022-07 to 2023-01 & 2.17 & 28.47 & 91.46 & 5.11 & 327.50 \\
        2023-02 to 2023-04 & 1.33 & 55.39 & 117.89 & 0.01 & 164.60 \\
        2023-05 to 2023-10 & 2.60 & 28.26 & 120.44 & 2.58 & 287.48 \\
        \hline
    \end{tabular}
    \caption{Duration of quasi-equilibrium states along with parameters of $p(S,H)$ in Eq. \ref{Eq:pSH}.}
    \label{tab:p(S,H)}
\end{table} 

Note that the identification of the quasi-equilibrium states depends on the considered time-scales. For example, by considering a larger time-scale we could have grouped together all of the data on Fig. \ref{fig:p(S,H)_lines} into a single statistical ensemble described by parameters $a = 2.22$, $b = 26.33$, $c = 88.78$, $\sigma_\perp^2 = 79.02$ and $\sigma_\parallel^2 = 912.70$, but then we would not be able to identify the more fine-grained details of the ``minor transitions in evolution'' discussed here.

Returning to the phase transitions between quasi-equilibrium states, consider the period from July 2021 to April 2022, as plotted in Fig. \ref{fig:p(S,H)_phase_trans}.
\begin{figure}
    \centering
    \includegraphics[width=1.0\textwidth]{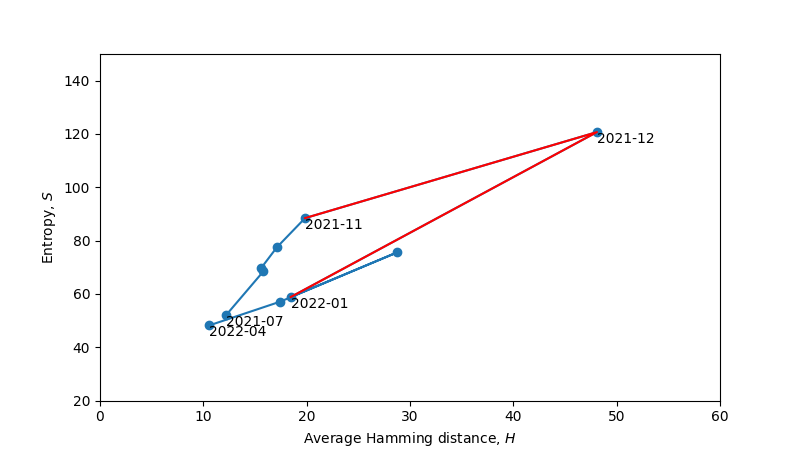}
    \caption{Scatter plot of the total entropy and average Hamming distance from July of 2021 to April of 2022. Quasi-equilibrium states are marked in blue and phase transition is marked in red.}
    \label{fig:p(S,H)_phase_trans}
\end{figure}
The two connected blue lines represent the two quasi-equilibrium states, and the red line indicates a single phase transition that occurred around December 2021. Other phase transitions (and quasi-equilibrium states) can be seen on Fig. \ref{fig:p(S,H)_all_trans}
\begin{figure}
    \centering
    \includegraphics[width=1.0\textwidth]{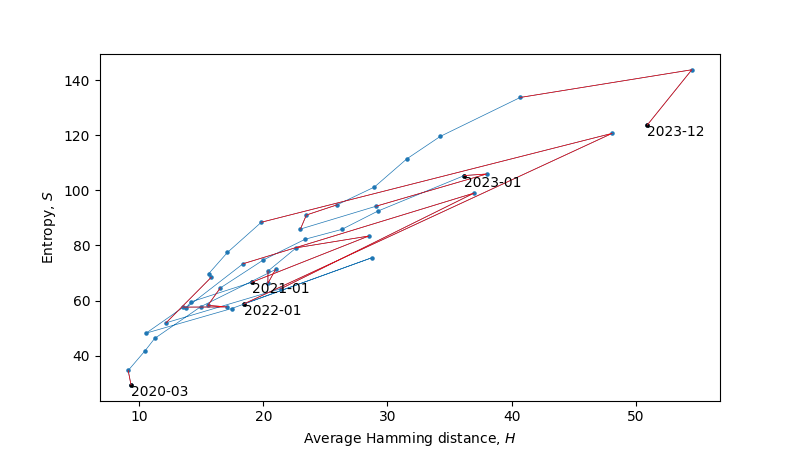}
    \caption{Scatter plot of the total entropy and average Hamming distances. Quasi-equilibrium states are marked in blue and phase transition are marked in red.}
    \label{fig:p(S,H)_all_trans}
\end{figure}
which is the same as Fig. \ref{fig:p(S,H)_lines} but with quasi-equilibrium states marked in blue and phase transitions marked in red. It appears that the system has recently undergone a phase transition and is currently transitioning to a new quasi-equilibrium state, the statistical properties of which are still to be revealed.

\section{Conclusion}\label{Sec:Conclusion}

The evolution of any biological system can be analyzed either microscopically (e.g., by tracking individual mutations) or macroscopically (e.g., by monitoring the dynamics of macroscopic or thermodynamic parameters). In this paper, we primarily focus on the macroscopic modeling of the total Shannon entropy (see Sec \ref{Sec:Entropy}) and average Hamming distance (see Sec. \ref{Sec:Hamming}) to study the evolution of the coronavirus using publicly available data from the United Kingdom between March 2020 and December 2023. In particular, we identify the so-called quasi-equilibrium states, when an approximate linear dependence between the total Shannon entropy and the average Hamming distance holds, and phase transitions between such states, when the linear dependence breaks down (see Sec. \ref{Sec:Equilibrium}).

The quasi-equilibrium state in the early pandemic corresponds to the prevalence of a single variant. However, in the late pandemic, the quasi-equilibrium states can acquire additional scales (or levels) and simultaneously describe multiple variants, forming what we call a multi-level quasi-equilibrium state. The numerical analysis suggests that the system is about to complete a phase transition to a new (and perhaps multi-level) quasi-equilibrium state, whose statistical properties are yet to be uncovered (see Sec. \ref{Sec:Equilibrium}).

In a quasi-equilibrium state, an evolving system constantly undergoes neutral mutations, exploring the neutral network. However, these mutations do not significantly increase or decrease its fitness. For the coronavirus, this typically lasts for a period of a few months until the virus undergoes an adaptive mutation that increases its fitness, providing it with an advantage over the old variants. This is a phase transition during which a statistical ensemble of viruses (and the corresponding macroscopic parameters such as Shannon entropy and Hamming distance) changes discontinuously which allowed us to consistently identify and study such phase transitions. 

The performed studies open a gateway for modeling evolutionary dynamics in terms of macroscopic parameters which need not be confined to only Shannon entropy and Hamming distances. As was discussed in the paper, for each quasi-equilibrium state there is a network of neutral mutations and statistical properties of the network can be analysed using either numerical or analytical tools such as partition functions, perturbative calculations, spectral methods, etc. For example, some of the quasi-equilibrium states contain central sequences which is evident from the analysis of single-sequence average Hamming distances (see Sec. \ref{Sec:Hamming}), but, in general, a lot more statistical and spectral information about the neutral networks remain to be uncovered. 

Furthermore, the conducted analysis has revealed the potential utility of statistical methods in establishing an early warning system for pandemics, offering independent value for humanity. The argument put forth suggests that a discontinuous change in the average Hamming distance or total Shannon entropy serves as a robust early indicator of a phase transition. While this concept makes sense, there might be an even more effective approach to detecting potentially perilous phase transitions. For instance, even if the total entropy remains continuous but experiences rapid growth, the effective space of possible mutations expands rapidly, thereby increasing the probability of an adaptive mutation. Utilizing the spectrum of the neutral network could enable predictions regarding when such mutations and subsequent phase transitions might occur. However, for making accurate predictions, additional spectral and statistical analyses are required, which we defer to future research.

\bibliographystyle{unsrt}
\bibliography{library}
\end{document}